\documentclass[superscriptaddress,twocolumn,showpacs,preprintnumbers,amsmath,amssymb,prb]{revtex4}

\usepackage{graphicx}
\usepackage{dcolumn}
\usepackage{bm}
\usepackage{multirow}
\usepackage{color}

\begin{document}


\title{Block magnetism coupled with local distortion in the iron-based spin-ladder BaFe$_2$Se$_3$}

\author{Yusuke Nambu}
\affiliation{Neutron Science Laboratory, Institute for Solid State Physics, University of Tokyo, 106-1 Shirakata, Tokai, Ibaraki 319-1106, Japan}
\affiliation{JST, TRIP, 5 Sanbancho, Chiyoda, Tokyo 102-0075, Japan}
\author{Kenya Ohgushi}
\affiliation{Institute for Solid State Physics, University of Tokyo, Kashiwanoha, Kashiwa, Chiba 277-8581, Japan}
\affiliation{JST, TRIP, 5 Sanbancho, Chiyoda, Tokyo 102-0075, Japan}
\author{Shunpei Suzuki}
\affiliation{Institute for Solid State Physics, University of Tokyo, Kashiwanoha, Kashiwa, Chiba 277-8581, Japan}
\author{Fei Du}
\affiliation{Institute for Solid State Physics, University of Tokyo, Kashiwanoha, Kashiwa, Chiba 277-8581, Japan}
\affiliation{College of Physics, State Key Laboratory of Superhard Materials, Jilin University, Changchun, China}
\author{Maxim Avdeev}
\affiliation{Bragg Institute, Australian Nuclear Science and Technology Organisation, PMB1, Menai, NSW 2234, Australia}
\author{Yoshiya Uwatoko}
\affiliation{Institute for Solid State Physics, University of Tokyo, Kashiwanoha, Kashiwa, Chiba 277-8581, Japan}
\affiliation{JST, TRIP, 5 Sanbancho, Chiyoda, Tokyo 102-0075, Japan}
\author{Koji Munakata}
\affiliation{Center for Neutron Science and Technology, CROSS, Tokai, Ibaraki 319-1106, Japan}
\author{Hiroshi Fukazawa}
\affiliation{Quantum Beam Science Directorate, Japan Atomic Energy Agency, Tokai, Ibaraki 319-1195, Japan}
\author{Songxue Chi}
\affiliation{Neutron Scattering Science Division, Oak Ridge National Laboratory, Oak Ridge, Tennessee 37831, USA}
\author{Yutaka Ueda}
\affiliation{Institute for Solid State Physics, University of Tokyo, Kashiwanoha, Kashiwa, Chiba 277-8581, Japan}
\affiliation{JST, TRIP, 5 Sanbancho, Chiyoda, Tokyo 102-0075, Japan}
\author{Taku J Sato}
\affiliation{Neutron Science Laboratory, Institute for Solid State Physics, University of Tokyo, 106-1 Shirakata, Tokai, Ibaraki 319-1106, Japan}
\affiliation{JST, TRIP, 5 Sanbancho, Chiyoda, Tokyo 102-0075, Japan}

\date{\today}

\begin{abstract}
Magnetism in the insulating BaFe$_2$Se$_3$ was examined through susceptibility, specific heat, resistivity and neutron diffraction measurements.
After formation of a short-range magnetic correlation, a long-range ordering was observed below $T_{\rm N}\sim 255$ K.
The transition is obscured by bulk properties.
Magnetic moments ($\parallel a$) are arranged to form a Fe$_4$ ferromagnetic unit, and each Fe$_4$ stacks antiferromagnetically.
This block magnetism is of the third type among magnetic structures of ferrous materials.
The magnetic ordering drives unusually large distortion via magnetoelastic coupling.
\end{abstract}


\maketitle

Since its discovery, research on iron-based superconductivity (SC) has become the main stream in condensed matter physics \cite{rev}.
The interplay between structure, magnetism and SC is one of most intriguing subjects of this field.
The common theme is on the 2D plane of iron, which has been realized in the ZrCuSiAs (1111), ThCr$_2$Si$_2$ (122), anti-PbO (11) and Cu$_2$Sb (111) structures.
To gain further insight into the mechanism of SC, realization of, for example, different crystal/magnetic structures is desired.
A recent prominent example is the 245 system \cite{245}, in which a novel type of magnetic structure in an insulator has been discovered.  
Regarding crystal structure, achievement of SC over various dimensions is also ambitious.
As spin ladders in copper oxides shed a new light on the mechanism of SC, a study on an analogue with ladder geometry among ferrous compounds is highly interesting.
Research on that compound can provide unique opportunity to study electronic correlations and importance of dimensionality for SC.

BaFe$_2$Se$_3$ is the first realization of ladder geometry among iron-based compounds \cite{bfs-chem}.
Unlike most of parent compounds of the iron-based superconductors, BaFe$_2$Se$_3$ is an insulator down to the lowest measured temperature.
The structure is novel, comprising of FeSe$_4$ tetrahedra with channels which host Ba atoms [Fig. 1(a)].
\begin{figure}[b!]
\includegraphics[width=240pt]{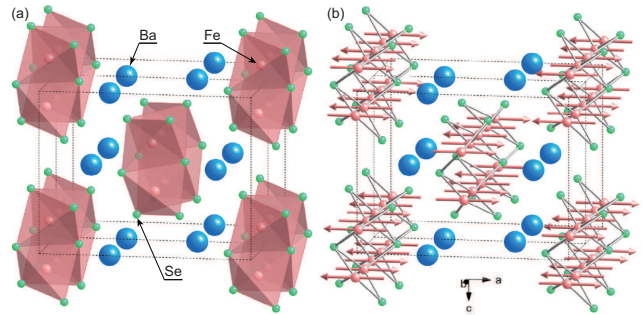}
\caption{(color online). (a) Crystal and (b) magnetic structures of BaFe$_2$Se$_3$ at $T=5$ K. Cuboids with dashed lines indicate crystallographic unit cells.}
\end{figure}
Four-fold coordinated Fe$^{2+}$ ions extend along the $b$-axis and forms two-leg spin-ladder structure within the $bc$-plane.

Here we report study of the first iron-based spin-ladder compound BaFe$_2$Se$_3$ based on high-quality polycrystalline sample.
High-resolution powder neutron diffraction reveals a 3D magnetic order below $T_{\rm N}\sim 255$ K.
However, bulk properties such as magnetic susceptibility, electrical resistivity and specific heat show no anomaly at the transition.
Magnetic moments ($\sim$ 2.8 $\mu_{\rm B}$ at 5 K) are arranged to form a Fe$_4$ ferromagnetic (FM) unit perpendicular to the ladder direction, and each Fe$_4$ stacks antiferromagnetically.
This block magnetism, as for in recently discovered 245 systems \cite{245}, is the realization of the third type of magnetic structures among iron-based materials.
Rietveld refinements on neutron diffraction data also reveal unusually large local lattice distortion driven by the magnetic order.

High-quality polycrystalline BaFe$_{2}$Se$_{3}$ was grown by the solid state reaction.
We first synthesized the precursor BaSe$_{3}$.
Stoichiometric amounts of elemental Ba shots and Se powder in an alumina crucible were sealed into an evacuated quartz ampoule.
It was slowly heated up to 700 $^{\circ}$C, kept for 12 hrs, and then quenched to room temperature. 
The obtained BaSe$_{3}$ and Fe with a 1:2 molar ratio were mixed together and put into an alumina crucible inside a quartz tube.
It was reacted at 700 $^{\circ}$C for 96 hrs with an intermediate regrinding.
The powder x-ray diffraction using Cu-$K_{\alpha}$ radiation indicates no trace of impurity phases, however, it turned out that there is a tiny amount of Fe$_{7}$Se$_{8}$ impurity by magnetization measurements.
The electrical resistivity was measured by the standard four-probe method. 
The magnetic data were collected by using a SQUID magnetometer.
The specific heat was measured by utilizing a commercial set-up (Quantum Design, PPMS), where the relaxation method is used.
Neutron powder diffraction data were collected on high-resolution ECHIDNA diffractometer at ANSTO and WAND at ORNL, with $\lambda=2.4395$ and 1.4827 \AA, respectively.
Diffraction patterns were obtained between $T=5$ and 350 K in a closed-cycle-refrigerator.
We employed group theoretical analysis to identify magnetic structures that are allowed by symmetry.

Table \ref{atomic} summarizes the structural parameters at $T=300$ K determined by Rietveld refinement on powder neutron diffraction data.
\begin{table}[tb]
\caption{Atomic positions within $Pnma$ of BaFe$_2$Se$_3$ at $T=300$ K determined by Rietveld analysis ($\chi^2=2.3$). Lattice constants are $a=11.93178(15)$ \AA, $b=5.43747(6)$ \AA, and $c=9.16476(12)$ \AA. Isotropic Debye-Waller factor ($U_{\rm iso}$) is employed.}
\label{atomic}
\begin{ruledtabular}
\begin{tabular}{lccccc}
Atom	&	Site	& $x$		& $y$		& $z$				& $U_{\rm iso}$ ({\AA}$^2$)	\\
\hline
Ba		& $4c$	& 0.1862(5)	& $1/4$		& 0.5154(10)	& 0.051(5)	\\
Fe		& $8d$	& 0.4941(3)	& 0.0000(6)	& 0.3525(3)	& 0.044(2)	\\
Se1	& $4c$	& 0.3595(4)	& $1/4$		& 0.2300(5)	& 0.059(5)	\\
Se2	& $4c$	& 0.6283(3)	& $1/4$		& 0.4923(6)	& 0.034(4)	\\
Se3	& $4c$	& 0.3997(4)	& $1/4$		& 0.8155(5)	& 0.061(5)	\\
\end{tabular}
\end{ruledtabular}
\end{table}
Consistent with other reports \cite{bfs1,bfs2,bfs3,bfs4}, the crystal structure is accounted for by orthorhombic space group $Pnma$.
By keeping site occupancy variable for each atoms in the refinement, the result indicates stoichiometric ratio within error bars, Ba$_{0.96(4)}$Fe$_{2.01(2)}$Se$_{3.02(7)}$. 
Temperature dependences of lattice parameters are provided in Fig. 2(a), showing systematic change throughout temperature regime.
\begin{figure}[tb]
\includegraphics[width=240pt]{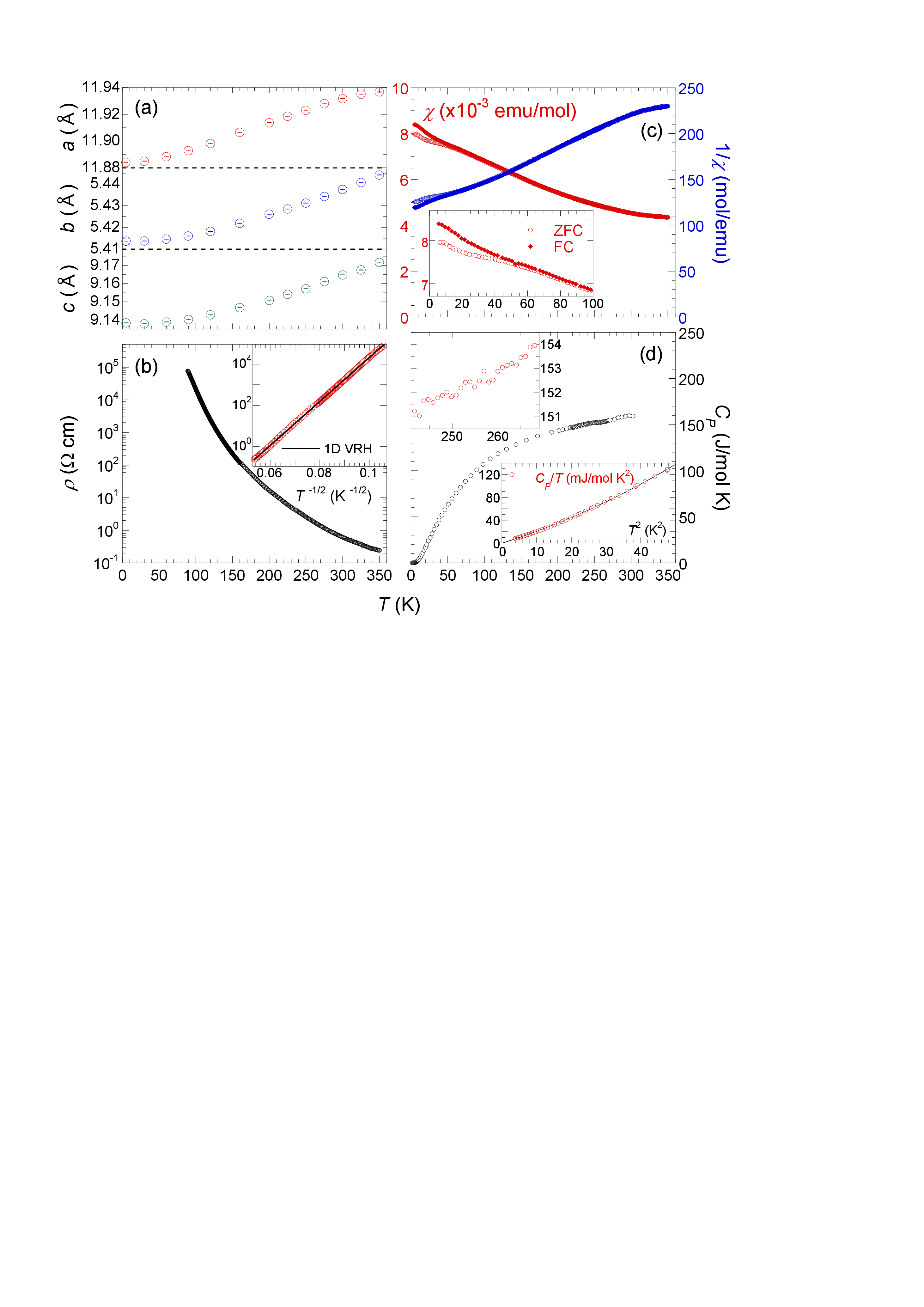}
\caption{(color online). Temperature dependences of (a) lattice parameters, (b) electrical resistivity ($\rho$), (c) magnetic susceptibility ($\chi$) and its inverse, and (d) specific heat ($C_P$). The insets of (b,c,d) show a fit with 1D variable-range-hopping model (see text), enlargement of bifurcation, anomaly around $T_{\rm N}$ (left top) and $C_P/T$ against $T^2$ with a polynomial fit (right bottom), respectively.}
\end{figure}

Electrical resistivity $\rho$ shows an insulating behavior [Fig. 2(b)].
Note that $\rho$ exhibits no anomaly in the whole temperature range measured.
A fitting with an activation-type formula $\rho = A\exp(\Delta/T)$ yields a small activation energy $\Delta=0.13$ eV, comparable with Ref \cite{bfs4}.
More precisely, $\rho$ deviates from the activation-type behavior, instead it shows a variable-range-hopping (VRH) type behavior.
The data is fit to $\rho\propto\exp(B+(\frac{D}{T})^{\frac{1}{1+d}})$, where $d$ denotes the dimension of the VRH model.
The VRH in 1D model yields better fit rather than higher dimensions [inset to Fig. 2(b)].
This indicates that the compound is classified as a Mott-Anderson insulator, and is in the vicinity of metal-insulator (MI) transition.
To further examine the conduction behavior, measurements on a single crystal are now under way.

Accurate magnetization measurement under small applied field is limited by a tiny amount of FM impurity, Fe$_{7}$Se$_{8}$ ($T_{\rm C}\sim 460$ K \cite{Fe7Se8}).
We therefore deduce the susceptibility $\chi\equiv M/H$ under $\mu_0 H=5$ T and subtract the one at 1 T to minimize effects from Fe$_7$Se$_8$.
The key issue on the susceptibility is that it shows a weak temperature variation in the paramagnetic phase (above $T_{\rm N}$) and does not obey the simple Curie-Weiss law [Fig. 2(c)].
One possible explanation for this is that a simple localized spin picture in a Mott-Anderson insulating state is invalid in the present system, which is close to the MI transition as anticipated by the small charge gap.
Another possibility could be that the unquenched orbital angular momentum is present in Fe$^{2+}$ ions owing to the prominence of the spin-orbit coupling in an orbital degenerated state.
In contrast to the earlier report \cite{bfs2} indicating an emergence of weak ferromagnetism below $T_{\rm N}$, our high-field susceptibility data show no prominent anomaly over the whole temperature range.
This possibility is also excluded by our powder neutron data.
As inset to Fig. 2(c) enlarged, $\chi$ in zero-field-cooled (ZFC) and field-cooled (FC) sequences show small hysteresis.
This could reflect glassy behavior caused by small randomness, but the origin of it is still unclear.
From the data of $\rho$ [Fig. 2(b)] and $\chi$ [Fig. 2(c)], we exclude the possibility of SC behavior of this compound, distinguished from Ref \cite{bfs1}.

Specific heat $C_P$ shows no anomaly in the whole temperature measured [Fig. 2(d)].
Given no peak formation at around $T_{\rm N}$ [left-top inset to Fig. 2(d)], entropy release associated with the magnetic ordering turned out to be surprisingly small.
A possible reason for this is that entropy release gradually occurs in a wide temperature range due to large fluctuations characterized in a low-dimensional system as will be discussed later.
$C_P/T$ versus $T^{2}$ plot at low temperatures indicates a negligible small $T$-linear contribution, as well as a small deviation from $C_P\propto T^3$ behavior [right-bottom inset to Fig. 2(d)].
The lack of $\gamma T$ term is in agreement with the Mott insulating nature, and inadequate fit with only $\beta T^3$ implies an existence of soft phonons.
The data can be well reproduced by a function $C_P=\beta T^3+ \delta T^5$ with $\beta=1.89$ mJ/mol K$^4$ and $\delta=0.017$ mJ/mol K$^{6}$.
If it is assumed that all the components of the $\beta$ term come from the phonon contribution, using the formula $\beta=12\pi^4 NR/5\theta_{\rm D}^3$ ($N$ being the number of atoms in a formula unit, 6), the Debye temperature ($\theta_{\rm D}$) is estimated to be 183 K.
This value is considerably small and further supports the existence of soft phonons. 

To gain microscopic information of magnetism, we performed high-resolution powder neutron diffraction.
The obtained data is well fit based only on BaFe$_2$Se$_3$, and Fe$_7$Se$_8$ impurity has so small volume fraction ($\lesssim 0.3$ \%) that the secondary phase is not taken into account in the refinements.
Additional resolution-limited peaks associated with magnetic ordering were observed below 250 K data.
Figure 3 shows a diffraction pattern taken at 5 K along with the Rietveld refinement.
\begin{figure}[tb]
\includegraphics[width=250pt]{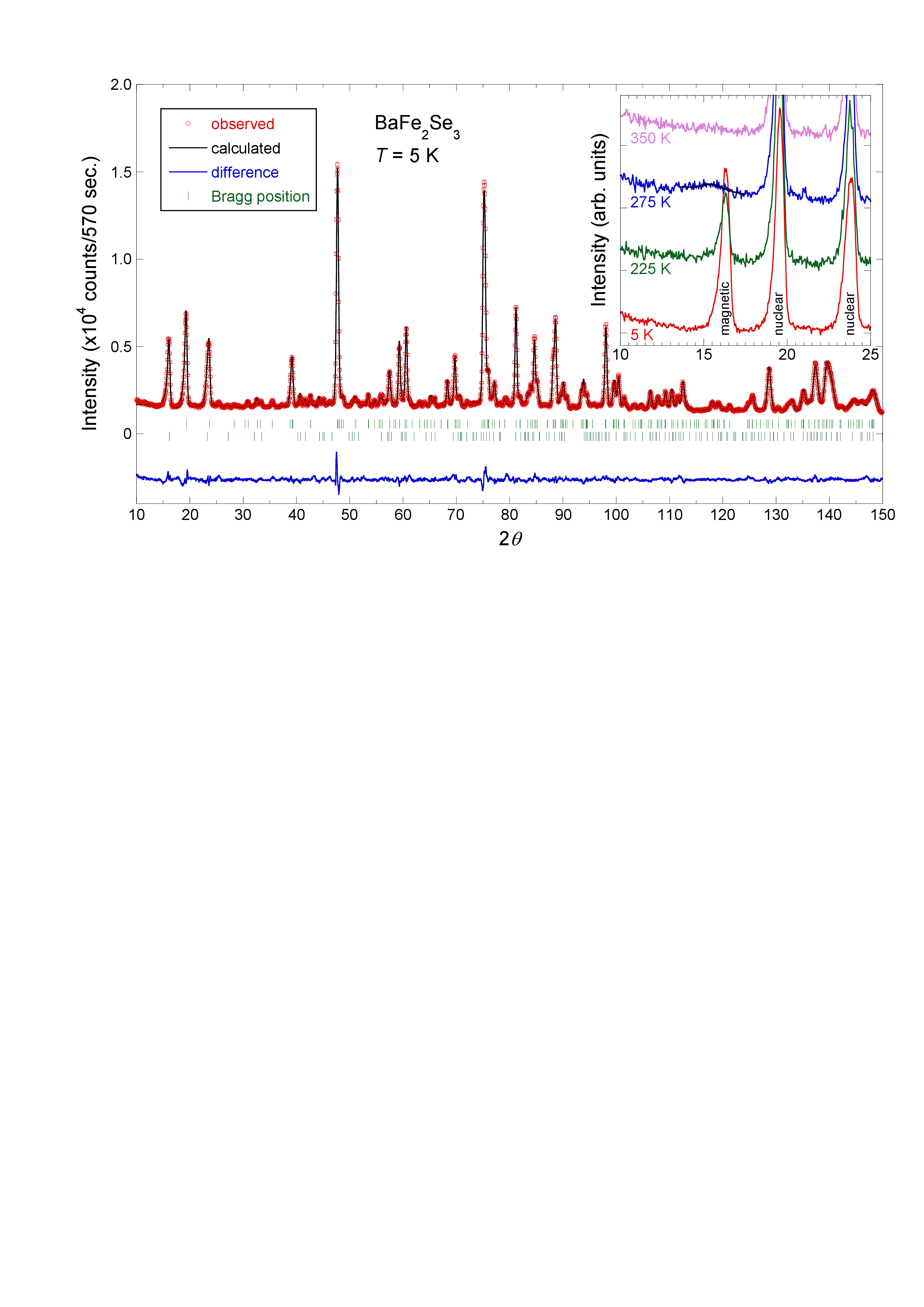}
\caption{(color online). (a) High-resolution neutron powder diffraction pattern of BaFe$_2$Se$_3$ at 5 K obtained on ECHIDNA with Rietveld refinement (solid lines). The calculated positions of nuclear and magnetic reflections are indicated (green ticks). The bottom line gives the difference between observed and calculated intensities. The inset shows temperature evolution of a magnetic reflection with short- (with black curve fit) and long-range correlations.}
\end{figure}
To identify the magnetic structure, we applied representation analysis.
All the magnetic peak positions can be accounted for by a propagation wave vector, $(1/2, 1/2, 1/2)$.
Basis vectors (BVs) of the irreducible representations (irreps) for the wave vector (Table \ref{irrep}) were obtained using the SARA{\it h} code \cite{SARAh}.
\begin{table}[tb]
\caption{Basis vectors (BVs) of irreducible representations (irreps) for the space group $Pnma$ with the magnetic wave vector ${\bm q}_{\rm m}=(1/2,1/2,1/2)$. Superscripts show the moment direction. Columns for positions represent \#1: $(x,y,z)$, \#2: $(x+1/2,-y+3/2, -z+1/2)$, \#3: $(-x+1,y-1/2,-z+1)$, \#4: $(-x+1/2,-y+1,z+1/2)$, \#5: $(-x+1,-y+1,-z+1)$, \#6: $(-x+1/2,y-1/2,z+1/2)$, \#7: $(x,-y+3/2,z)$, and \#8: $(x+1/2,y,-z+1/2)$.}
\label{irrep}
\begin{ruledtabular}
\begin{tabular}{cccccccccc}
irrep	&	BV	& \#1	& \#2	& \#3	& \#4	& \#5	& \#6	& \#7	& \#8	\\
\hline
\multirow{12}{*}{$\Gamma_1$}	& $\psi_1$	&	$2^a$	& 0	& 0	& 0	& 0	& 0	& $-2^a$	& 0 \\
	& $\psi_2$	&	$2^b$	& 0	& 0	& 0	& 0	& 0	& $2^b$	& 0 \\
	& $\psi_3$	&	$2^c$	& 0	& 0	& 0	& 0	& 0	& $-2^c$	& 0 \\
	& $\psi_4$	&	0	& 0	& 0	& $2^a$	& 0	& $2^a$	& 0	& 0 \\
	& $\psi_5$	&	0	& 0	& 0	& $2^b$	& 0	& $-2^b$	& 0	& 0 \\
	& $\psi_6$	&	0	& 0	& 0	& $-2^c$	& 0	& $-2^c$	& 0	& 0 \\
	& $\psi_7$	&	0	& 0	& 0	& $-2^a$	& 0	& $2^a$	& 0	& 0 \\
	& $\psi_8$	&	0	& 0	& 0	& $-2^b$	& 0	& $-2^b$	& 0	& 0 \\
	& $\psi_9$	&	0	& 0	& 0	& $2^c$	& 0	& $-2^c$	& 0	& 0 \\
	& $\psi_{10}$	&	$2^a$	& 0	& 0	& 0	& 0	& 0	& $2^a$	& 0 \\
	& $\psi_{11}$	&	$2^b$	& 0	& 0	& 0	& 0	& 0	& $-2^b$	& 0 \\
	& $\psi_{12}$	&	$2^c$	& 0	& 0	& 0	& 0	& 0	& $2^c$	& 0 \\
\hline
\multirow{12}{*}{$\Gamma_2$}	& $\psi_{13}$	&	0	& $-2^a$	& 0	& 0	& 0	& 0	& 0	& $-2^a$ \\
	& $\psi_{14}$	&	0	& $2^b$	& 0	& 0	& 0	& 0	& 0	& $-2^b$ \\
	& $\psi_{15}$	&	0	& $2^c$	& 0	& 0	& 0	& 0	& 0	& $2^c$ \\
	& $\psi_{16}$	&	0	& 0	& $2^a$	& 0	& $-2^a$	& 0	& 0	& 0 \\
	& $\psi_{17}$	&	0	& 0	& $-2^b$	& 0	& $-2^b$	& 0	& 0	& 0 \\
	& $\psi_{18}$	&	0	& 0	& $2^c$	& 0	& $-2^c$	& 0	& 0	& 0 \\
	& $\psi_{19}$	&	0	& 0	& $2^a$	& 0	& $2^a$	& 0	& 0	& 0 \\
	& $\psi_{20}$	&	0	& 0	& $-2^b$	& 0	& $2^b$	& 0	& 0	& 0 \\
	& $\psi_{21}$	&	0	& 0	& $2^c$	& 0	& $2^c$	& 0	& 0	& 0 \\
	& $\psi_{22}$	&	0	& $2^a$	& 0	& 0	& 0	& 0	& 0	& $-2^a$ \\
	& $\psi_{23}$	&	0	& $-2^b$	& 0	& 0	& 0	& 0	& 0	& $-2^b$ \\
	& $\psi_{24}$	&	0	& $-2^c$	& 0	& 0	& 0	& 0	& 0	& $2^c$ \\
\end{tabular}
\end{ruledtabular}
\end{table}
There are 24 BVs in total, belonging to distinct two irreps $\Gamma_1$ and $\Gamma_2$.
Each irrep has 12 BVs, in which only 4 of 8 Fe atoms in a unit cell are allowed to possess magnetic moments, and every BV describes the relation of moment direction of 2 atoms either parallel or antiparallel along one crystallographic axis.
Both irreps in this case are required to participate as a corepresentation to let all 8 atoms have magnetic moments.
In a certain moment direction along either the $a$, $b$ or $c$-axis, there are 2 choices for 4 combinations, namely 16 patterns. 
We assumed that the moment is parallel or antiparallel to one axis and has the same coefficient of 4 BVs, and sort out all 48 patterns to determine the magnetic structure by comparing $R$-factors.
The best fit with $R_{\rm mag}=8.54$ \% is a combination of BVs, $\psi_4$, $\psi_{10}$, $\psi_{16}$ and $\psi_{22}$, which is consistent with Ref \cite{bfs2}.
The estimated moment at 5 K is 2.75(2) $\mu_{\rm B}$ per an Fe$^{2+}$ site.
The 5 K magnetic structure corresponding to the best fit is shown in Fig. 1(b).
Magnetic moments ($\parallel a$) are arranged to form a Fe$_4$ FM unit in the $bc$-plane, and each Fe$_4$ stacks antiferromagnetically.
For inter-ladder coupling on the $ac$-plane, block spins that are AF correlated in one neighboring ladder and ferromagnetically correlated in the perpendicular direction.
The next nearest-neighbor (nn) ladders interact antiferromagnetically.
The determined magnetic structure, described as ``block magnetism'', is reminiscent of one observed in insulating $A_2$Fe$_4$Se$_5$ ($A=$ K, Rb, Cs, Tl) \cite{245} with vacancy-ordered 2D Fe plane.

Despite diverse structural and physical properties, there have been so far reported only several types of magnetic structures among iron-based materials.
The common theme of magnetism and SC is on a square lattice of Fe with either pnictogen or chalcogen.
In cuprates SC, parent compounds as Mott insulators show a simple AF pattern, so-called checkerboard magnetic structure.
This is favored by strong AF nn interactions, yet has not been observed in any ferrous compounds.
The most common spin structure among the ferrous is single-stripe magnetic structure.
This features AF nn spins in one direction and ferromagnetically correlated in the perpendicular direction.
Stripes extend along nn bonds as favored by next nn AF exchange interactions.
The single-stripe is observed in the parent compounds of the 1111, 122 and 111 materials.
Double-stripe is also widely observed in the 11 systems \cite{Bao} and insulating La$_2$O$_2$Fe$_2$OSe$_2$ \cite{La2O2Fe2OSe2}, in which stripes extending again along nn bonds in a structure that is favored by AF third nn interactions.
The magnet CaFe$_4$As$_3$ with similar FeAs strips but connected in 3D has an incommensurate variant of the nearly double-stripe structure \cite{cfs1,cfs2}.
Finally, block magnetism has been recognized since the discovery of the 245 systems, in which nn 4 Fe atoms connected FM and Fe plane is occupied by that block interacting AF with a separation of Fe-vacancy.
Note that this structure has magnetic moments perpendicular to the Fe plane, whereas the other structures have within plane moments.
There is a close relation between the magnetic structure and moment size; single-stripe typically has typically $\lesssim$ 1 $\mu_{\rm B}$, double-stripe $\gtrsim$ 2 $\mu_{\rm B}$ and block magnetism around 3 $\mu_{\rm B}$.
This classification is also the case for BaFe$_2$Se$_3$, which has large ($\sim 2.8$ $\mu_{\rm B}$) magnetic moments perpendicular to the ladder-plane, belonging to the block magnetism.

Temperature dependence of the determined magnetic moment is shown in Fig. 4(b).
\begin{figure}[tb]
\includegraphics[width=220pt]{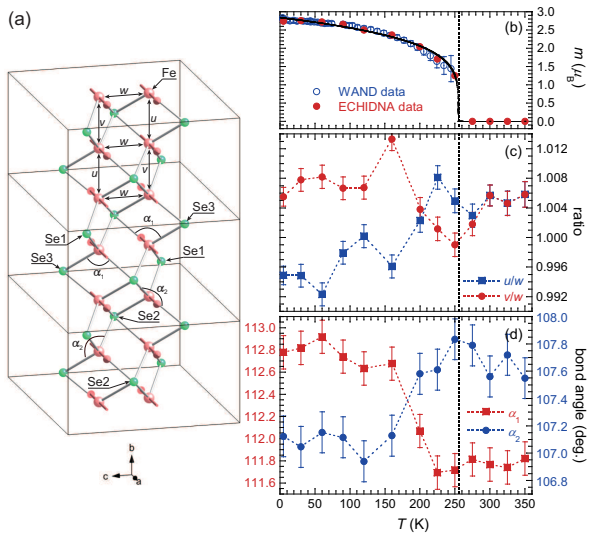}
\caption{(color online). (a) A ladder geometry in BaFe$_2$Se$_3$. Solid lines surround unit cells. Temperature dependence of (b) estimated magnetic moment of the LRO, (c) ratio between lengths along the ladder ($u,v$) and leg ($w$) directions, and (d) representative angles ($\alpha_1, \alpha_2$) of Se-Fe-Se.}
\end{figure}
Together with ECHIDNA and WAND data, $T_{\rm N}$ is roughly estimated to be $254.6\pm 2.4$ K by an order parameter-type fit.
However, previous M\"ossbauer experiment \cite{bfs-Moessbauer} reports no anomaly at around $T_{\rm N}$, instead clarifies the appearance of magnetic hyperfine splitting below 100 K.
This distinct behavior could originate from the different timescale of the experimental technique; neutron typically has 10$^{-12}$ sec timescale, being faster than that of M\"ossbauer (10$^{-7}$ sec).
Such a gradual slowing down of spin fluctuations is sometimes observed in low-dimensional magnets.
If entropy release between 100 K and $T_{\rm N}$ was sufficient, the obscurity of it at the LRO might be uncontradicted.
Moreover, we observed a broad magnetic peak above $T_{\rm N}$ at slightly off position ($Q \sim 0.68$ {\AA}$^{-1}$) from 0.73 {\AA}$^{-1}$ of the LRO [inset to Fig. 3].
This diffusive peak with the estimated correlation length of 19(5) {\AA}, could also reflect the low-dimensionality of the material.
Elucidation of spin dynamics and short-range correlation realized by BaFe$_2$Se$_3$ would be important, and will be done shortly.

Despite the systematic lattice shrinkage as temperature decreased [Fig. 2(a)], significant local lattice distortion driven by the magnetic order is observed.
Figure 4(a) depicts one ladder, where the summation of lengths, $u+v$ is equivalent to the lattice constant $b$, and $w$ is the width of the ladder.
Above $T_{\rm N}$, $u/w$ and $v/w$ is completely overlapping each other and takes slightly off value from the exact ladder (1.000) as shown in Fig. 4(c).
They start to deviate after the magnetic order occurs, $v/w$ shows a small increasing or constant behavior, whereas $u/w$ is surely decreasing on cooling.
This is attributed to change of the $y$ position of Fe.
We note that the atomic displacement along the $b$-axis of Fe below $T_{\rm N}$ is as large as approximately 0.01 {\AA}, which exceeds 10$^{-5}$ {\AA} reported in ordinary materials \cite{magneto-elastic}.
This is in contrast to the 245 systems which show a shrinkage to the center of Fe$_4$ direction as favored by nn FM interactions \cite{block}.
The distortion in BaFe$_2$Se$_3$ is also evidenced by change of Se-Fe-Se bond angles [Fig. 4(d)].
Even above $T_{\rm N}$, an FeSe$_4$ tetrahedron is distorted, and the magnetic ordering enhances it by elongating the lattice along the $c$.
This splits $e$ orbitals and stabilizes $x^2-y^2$ against $3z^2-r^2$, where the direction of $z$ orbital is perpendicular to the ladder plane.
BaFe$_2$As$_2$ shows also similar temperature evolution of As-Fe-As angles across tetragonal to orthorhombic structural transition \cite{Ba122}.
However, the angle change in BaFe$_2$As$_2$ is half as large as in BaFe$_2$Se$_3$.
Given these facts of significant lattice distortion inherent, the $x^2-y^2$ orbital becomes firmly stabilized below $T_{\rm N}$.
Again, no symmetry change occurs at all in the temperature range from 5 to 350 K in BaFe$_2$Se$_3$.
The orthorhombic BaFe$_2$Se$_3$ undergoes such a strong change due to magnetoelastic coupling without breaking the high-temperature symmetry.
Instead it gives rise to the unusually large atomic displacement of Fe within the unit cell, and it can be consistent with soft-phonon mode inferred from the specific heat results.
It has been reported the orthorhombicity ratio in BaFe$_2$As$_2$ is up to $3\times 10^{-3}$, and it is systematically suppressed by electron doping to finally reach to SC \cite{orthorhombicity}.
BaFe$_2$Se$_3$, on the other hand, has almost twice large distortion.
Since this type of distortion is not preferred by SC in iron-based materials, applying pressure and doping studies are now under way to accomplish SC.

In summary we have examined Fe magnetism in the spin-ladder geometry realized in BaFe$_2$Se$_3$.
A magnetic short-range correlation is detected, followed by a 3D long-range ordering.
The transition is, however, obscured by bulk properties measurements.
The determined magnetic structure is so-called block magnetism, where $\parallel a$ moments are arranged to form a Fe$_4$ FM unit, and each Fe$_4$ stacks antiferromagnetically.
The magnetic ordering also drives unusually large lattice distortion, which SC in ferrous compounds does not prefer.

We thank M. Isobe for technical assistance and K. Okazaki for discussion.
The work is supported in part by Grant-in-Aids for Scientific Research (2340097) and the Japanese Society for Neutron Science.

\end{document}